\documentstyle[12pt]{article}
\newcommand{\be}{\begin{equation}}
\newcommand{\ee}{\end{equation}}
\newcommand{\ba}{\begin{array}{l}}
\newcommand{\ea}{\end{array}}

\newcommand{\ci}[1]{\cite{#1}}
\newcommand{\banonum}{\begin{eqnarray*}}
\newcommand{\eanonum}{\end{eqnarray*}}
\newcommand{\baa}{\begin{eqnarray}}
\newcommand{\eaa}{\end{eqnarray}}
\newcommand{\bfr}{\begin{flushright}}
\newcommand{\efr}{\end{flushright}}
\newcommand{\bfl}{\begin{flushleft}}
\newcommand{\efl}{\end{flushleft}}
\begin{document}
\date{}
\title{Spectra of heavy-light mesons at finite temperature}
\author{D.U.~Matrasulov\thanks{E-mail: dmatrasu@phys.ualberta.ca},
Kh.T.~Butanov, Kh.Yu.~Rakhimov\\
\it{Heat Physics Department of the Uzbek Academy of Sciences,}\\
\it{28 Katartal St., 100135 Tashkent, Uzbekistan}\\
and\\
F.C.~Khanna\\
\it{Physics Department University of Alberta}\\
\it{Edmonton Alberta, T6G 2J1 Canada}\\}

\maketitle
\begin{abstract}
Spectra of heavy-light meson are studied using potential model and thermofield dynamics prescription. The mass spectra of different heavy-light mesons are calculated at different temperatures and compared with those at $T=0$. It is found that the binding mass of heavy-light meson decreases as temperature increases.
\end{abstract}

\section{Introduction}

Recent experiments on heavy-light mesons at BABAR~\ci{babar}, CLEO~\ci{cleo} and BELLE~\ci{belle}
collaborations stimulated extensive theoretical studies.
Within QCD, relativistic potential models, chiral perturbation theory
and non-relativistic potential models that are analogs of the
hydrogen-like atoms. All numerical work is based on lattice QCD
gives probably the best results \ci{qcd1}-\ci{lattice}.

An effective method to explore hadron properties at finite
temperatures is the lattice QCD approach that leads to
results of higher precision for spectra and transition
rates. However it requires powerful computer resources. It is
therefore interesting to explore methods that are simpler
and yield interesting results for the hadron spectra at
finite temperature. Recently such a method has been used to
calculate quarkonium spectra at finite temperature in
potential models \ci{matr1}. In this paper we extend this method to
heavy-light mesons with its dynamics described by the Dirac
equation. The finite temperature effects are included by
using thermofield dynamics (TFD), a real time finite
temperature field theory. We apply the prescription to the
Dirac Hamiltonian with Coulomb and a linear potential. The
physical content of TFD will be described below. It is
important to note that TFD has been used to study the Casimir
effect and quantum chaos in Yang-Mills-Higgs system at
finite temperature~\cite{ndfi04}.

With the results emerging from RHIC (Relativistic Heavy
Ion Collider) and the anticipated results from LHC (Large
Hadron Collider) properties at finite temperature will be
needed. In the process of converting baryonic system to
quark-gluon plasma and subsequently the hadronization of
this plasma would have mesons and baryons at finite
temperature. This would require how these hadrons behave at
high temperatures. The question of reactions, change of
coupling constants and decay rates finite temperatures have
been considered already~\cite{rakhimov01}. The present calculation provides
additional information as to how the mass changes with
temperature.

The plan of the paper is to consider briefly the
description of the heavy-light mesons using the Dirac equation
at zero temperature. In section~\ref{tfd}, we give a short review of
TFD in order to make clear the finite temperature
formulation in general, stating clearly the underlying
details of the theory. In section~\ref{spectr}, TFD is applied to
calculate the spectra of heavy-light mesons at finite
temperature. Some concluding remarks are given in the final
section.

\section{Dirac equation for heavy-light meson}

Starting point for our study of finite-temperature mass spectra of heavy-light meson is
the Dirac equation written as \ci{dm,morii}
\be
i\frac{\partial \Psi}{\partial t}=H\Psi
\ee
where $H=\vec\alpha\vec p+\beta\left(m_q+\lambda r\right)-\frac{Z}{r}+V_0$.
We note that the confining part of the potential is included into this
equation as a Lorentz scalar (not as fourth component of the 4-vector).
Introducing the following time scaling $\tau = r^{-1}t$, we can rewrite this
equation as
\be
i\frac{\partial\Psi}{\partial\tau}=rH\Psi.
\ee
Using standard substitution
$\Psi(r,\theta,\tau)=e^{-iE\tau}\Psi(r,\theta)$ and separating angular
variables we get the time-independent radial equation:
\be
E\Psi=rH_r\Psi,
\ee
where
\be
H_r=\left(\begin{array}{cc}m_q+\lambda r+V_0-\frac{Z}{r}&-\frac{d}{dr}+\frac{\kappa}{r}\\\frac{d}{dr}+\frac{\kappa}{r}&-m_q-\lambda r-V_0-\frac{Z}{r}\end{array}\right)
\label{dir}
\ee
being the radial Dirac Hamiltonian.
For our purpose we need to represent this Hamiltonian in terms of annihilation and creation operators which are defined as
\be
a=\frac{1}{\sqrt{2}}\frac{d}{dr}+\frac{1}{\sqrt{2}}r, \;\;\;\;
a^\dagger=-\frac{1}{\sqrt{2}}\frac{d}{dr}+\frac{1}{\sqrt{2}}r.
\ee
It is easy to see that these operators obey the following commutation relation
$$
[a,a^\dagger] =1.
$$

The radial Dirac Hamiltonian~(\ref{dir}) in terms of these operators can be written as
\be
H_r=r^{-1}\left(\begin{array}{cc}A&B\\C&D\end{array}\right)\label{rdirac}
\ee
where
$$
A=\frac{m_q+V_0}{\sqrt{2}}(a+a^\dagger)+\frac{\lambda}{2}(a+a^\dagger)^2-Z,
$$
$$
B=-\frac{1}{2}(a+a^\dagger)(a-a^\dagger)+\kappa,
$$
$$
C=\frac{1}{2}(a+a^\dagger)(a-a^\dagger)+\kappa,
$$
and
$$
D=-\frac{m_q+V_0}{\sqrt{2}}(a+a^\dagger)-\frac{\lambda}{2}(a+a^\dagger)^2-Z.
$$
The eigenvalues of the Hamiltonian written in terms of annihilation and creation operators can be calculated by diagonalizing
it in a harmonic oscillator basis. The wave function is expanded in terms of nonrelativistic harmonic oscillator
wave functions as
\be
\mid \Psi\rangle=\sum\limits_n\left(\begin{array}{c}a_n\\b_n\end{array}\right)\mid n\rangle.
\ee
Then the energy eigenvalues of the heavy-light meson are to be calculated by diagonalization of the matrix
$$
E=\langle\Psi'\mid rH_r\mid\Psi\rangle.
$$

This method will be combined with the thermofield dynamics to calculate spectra of heavy-light meson at finite temperature.
The next section presents briefly basic thermofield dynamics.

\section{Thermofield Dynamics}\label{tfd}

Thermofield dynamics is a real-time operator formalism of quantum
field theory at finite temperature. It has been recognized that
for a theory at finite temperature, the Hilbert space has to be
doubled~\cite{tak,das}. The usual Hilbert space may be defined by
creation, $a^{\dagger}(k)$ and annihilation operators, $a(k)$.
Then the second Hilbert space is defined by tilde operators,
$\tilde a^{\dagger}(k)$ for creation and $\tilde a(k)$ for
annihilation operators. These operators, for a boson system,
satisfy
$$
[a(k),a^{\dagger}(k')]=\delta_{k,k'},
$$
$$
[\tilde a(k),\tilde a^{\dagger}(k')]=\delta_{k,k'},
$$
$$
[a(k),\tilde a^{\dagger}(k')]=0
$$
and all other commutators are zero. For fermion systems anti-commutator are to be used.

At finite temperature a Bogolyubov transformation is used to mix these two sets of operators to obtain the finite temperature creation and annihilation operators. Any operator $A(x)$ is then written at finite temperature as
$$
A(x,\theta)=U(\theta)\; A(x)\; U^{-1}(\theta)
$$
where $U(\theta)$ is the Bogolyubov transformation, i.e.
$$
U(\theta)=e^{-iG(\theta(\beta))}
$$
where $G=i\theta(\beta)(a^{\dagger}\tilde a^{\dagger}-a\tilde a)$ with $\beta=\frac{1}{k_{\rm B}T}$. Here $T$ is the temperature and $k_{\rm B}$ is the Boltzmann constant.

This implies that
$$
a(k,\theta)=U(\theta)a(k)U^{-1}(\theta)=u(\theta)a(k)-\vartheta(\theta)\tilde a^{\dagger}(k)
$$
$$
a^{\dagger}(k,\theta)=U(\theta)a^{\dagger}(k)U^{-1}(\theta)=u(\theta)a^{\dagger}(k)-\vartheta(\theta)\tilde a(k)
$$
$$
\tilde a(k,\theta)=U(\theta)\tilde a(k)U^{-1}(\theta)=u(\theta)\tilde a(k)-\vartheta(\theta)a^{\dagger}(k)
$$
$$
\tilde a^{\dagger}(k,\theta)=U(\theta)\tilde a^{\dagger}(k)U^{-1}(\theta)=u(\theta)\tilde a^{\dagger}(k)-\vartheta(\theta)a(k)
$$
where $u(\theta)=\cosh(\theta)$, $\vartheta(\theta)=\sinh(\theta)$ such that $u^2(\theta)-\vartheta^2(\theta)=1$. The vacuum state at finite temperature is defined similars as
$$
\mid {\rm 0}(\theta)\rangle=U(\theta)\mid \rm 0, \tilde 0\rangle.
$$
Then the temperature dependent creation and annihilation operators acting on the vacuum state, $\mid {\rm 0}(\theta)\rangle$ give
$$
a(k,\theta)\mid {\rm 0}(\theta)\rangle=\tilde a(k,\theta)\mid {\rm 0}(\theta)\rangle=0.
$$
Thus there is a procedure to work with the two set of operators
$$ (A_i\; A_j\tilde) = \tilde{A_i}\;\tilde{A_j} $$
$$ (cA_i + A_j\tilde) =c^*\tilde{A_i} + \tilde{A_j} $$
$$ (A_i^*\tilde) = (\tilde{A_i})^{\dagger} $$
$$ (\tilde{A_i}\tilde) = A_i $$
and $[\tilde{A_i},\; A_j]=0$. It is to be noted that $\vartheta^2(\theta)=\sinh^2\theta=[e^\beta-1]^{-1}$ where $\beta=\frac{\omega}{k_{\rm B}T}$.

From a group theoretical approach, it has been established that
for any kinematical symmetry there are two sets of operators,
physical observable $O$ and generators of symmetry, $\hat
O=O-\tilde O$. This establishes a connection to an algebraic
approach based on kinematic symmetry and the doubling does not
happen due to an arbitrary ansatz~\cite{Ademir}.

Such a procedure has been used to calculate scattering cross section and decay rates at finite temperature. The use of the Bogolyubov transformations that mixes the operators from the two Hilbert spaces is a reminder of superconductivity where this implied a condensation of quanta in the vacuum state.

Thus TFD is a powerful tool for exploring quantum dynamics of a
system at finite temperature provided its Hamiltonian can be
represented in terms of annihilation and creation operators. It
has found many applications in condensed matter physics
\ci{um,tadic,egor}, especially in superconductivity theory and
related topics. Recently TFD has been applied to explore quantum
chaos in the Yang-Mills-Higgs system \ci{matr} and for the
calculation of the spectra of a strongly interacting bound system
\ci{matr1}. In this work we apply the TFD prescription to the Jaynes-Cummings model. It should be noted that TFD has been
applied earlier to the Jaynes-Cummings model \ci{barnet} where the
thermal noise effects in quantum optics are studied.
In the next section we apply the TFD
prescription for the calculation of the energy spectra of heavy-light mesons.

\section{Spectra of heavy-light meson at finite temperature}\label{spectr}

Applying TFD prescription to the Hamiltonian of the heavy-light meson we can write it in temperature-dependent form.
Then finite-temperature energy eigenvalues of heavy-light meson can be calculated by diagonalization of the following temperature-dependent matrix:
$$
E(\beta)=\langle\Psi'_{\beta}\mid rH_r\mid\Psi_{\beta}\rangle=\left(\begin{array}{cc}a_{n',n}(\beta)&b_{n',n}(\beta)\\c_{n',n}(\beta)&d_{n',n}(\beta)\end{array}\right),
$$
where
$$
\mid \Psi_{\beta}\rangle=\sum\limits_n\left(\begin{array}{c}a_n\\b_n\end{array}\right)\mid n(\beta)\rangle,
$$

$$
a_{n',n}(\beta)=\frac{m_q+V_0}{\sqrt{2}}A_1(\beta)+\frac{\lambda}{2}A_2(\beta)-Z\,\delta_{n',n},
$$
$$
b_{n',n}(\beta)=-\frac{1}{2}A_3(\beta)+(\kappa+\frac{1}{2})\,\delta_{n',n},
$$
$$
c_{n',n}(\beta)=\frac{1}{2}A_3(\beta)+(\kappa-\frac{1}{2})\,\delta_{n',n},
$$
$$
d_{n',n}(\beta)=-\frac{m_q+V_0}{\sqrt{2}}A_1(\beta)-\frac{\lambda}{2}A_2(\beta)-Z\,\delta_{n',n},
$$
and
$$
A_1(\beta)=\left(\sqrt{n}\;\delta_{n',n-1}+\sqrt{n+1}\;\delta_{n',n+1}\right)(\cosh\theta+\sinh\theta),
$$
$$
A_2(\beta)=\left(\sqrt{n(n-1)}\;\delta_{n',n-2}+\sqrt{(n+1)(n+2)}\;\delta_{n',n+2}+(1+2n)\delta_{n',n}\right)(\cosh^2\theta+\sinh^2\theta)
$$
$$
+2\cosh\theta\sinh\theta((n+1)\,\delta_{n',n+1}+n\,\delta_{n',n-1}),
$$
$$
A_3(\beta)=\left(\sqrt{n(n-1)}\;\delta_{n',n-2}-\sqrt{(n+1)(n+2)}\;\delta_{n',n+2}\right)(\cosh^2\theta+\sinh^2\theta).
$$

Using this method we can calculate finite-temperature energy and mass spectra of heavy-light mesons.

\section{Results}

In table~\ref{tab1} mass spectra of $c\bar s$-meson is presented for different values of principle and
orbital quantum numbers at different temperatures.
The parameters are chosen as follows:
$m_{\bar s}=0.5$ GeV, $m_{c}=1.486$ GeV, $\lambda=0.8$ GeV$^2$, $V_0=-0.6$ GeV,
$\alpha_s=0.32$ ($Z=\frac{4}{3}\alpha_s$).
It is clear from this table that increasing the temperature
leads to decreasing of the mass. Such effect has been found earlier in the case of
quarkonium spectra at finite temperature \ci{matr1,chai}.
Table~\ref{tab2} represents the spectrum of $b\bar s$-system for the following set of parameters:
$m_{\bar s}=0.5$ GeV, $m_{b}=4.88$ GeV, $\lambda=0.8$ GeV$^2$, $V_0=-0.6$ GeV,
$\alpha_s=0.22$. Table~\ref{tab3} represents the spectrum of $b\bar c$-system for the following parameters:
$m_{\bar c}=1.486$ GeV, $m_{b}=4.88$ GeV, $\lambda=0.2$ GeV$^2$, $V_0=-0.3$ GeV,
$\alpha_s=0.22$.

\begin{figure}
\setlength{\unitlength}{0.6mm}\thicklines
\begin{center}
\begin{picture}(200,200)
\put(30,0){\framebox(180,160)}

\put(38,32.46){\line(1,0){2}}
\put(42,32.46){\line(1,0){2}}
\put(46,32.46){\line(1,0){2}}
\put(50,32.46){\line(1,0){2}}
\put(54,32.46){\line(1,0){2}}
\put(58,32.46){\line(1,0){2}}
\put(62,32.46){\line(1,0){2.3}}

\put(38,36.18){\line(1,0){4}}
\put(45.5,36.18){\line(1,0){4}}
\put(53,36.18){\line(1,0){4}}
\put(60.5,36.18){\line(1,0){4}}

\put(38,43.98){\line(1,0){9}}
\put(50,43.98){\line(1,0){2.2}}
\put(55.2,43.98){\line(1,0){9}}

\put(38,51.5){\line(1,0){26.5}}
\put(46,22.46){$1S$}

\put(80,63.14){\line(1,0){2}}
\put(84,63.14){\line(1,0){2}}
\put(89,63.14){\line(1,0){2}}
\put(93,63.14){\line(1,0){2}}
\put(97,63.14){\line(1,0){2}}
\put(101,63.14){\line(1,0){2}}
\put(105,63.14){\line(1,0){2.3}}

\put(80,66.72){\line(1,0){4}}
\put(87.5,66.72){\line(1,0){4}}
\put(95,66.72){\line(1,0){4}}
\put(102.5,66.72){\line(1,0){4}}

\put(80,74.6){\line(1,0){9}}
\put(92,74.6){\line(1,0){2.2}}
\put(97.2,74.6){\line(1,0){9}}

\put(80,89.24){\line(1,0){26.5}}
\put(91,53.14){$2S$}

\put(122,92.9){\line(1,0){2}}
\put(126,92.9){\line(1,0){2}}
\put(130,92.9){\line(1,0){2}}
\put(134,92.9){\line(1,0){2}}
\put(138,92.9){\line(1,0){2}}
\put(142,92.9){\line(1,0){2}}
\put(146,92.9){\line(1,0){2.3}}

\put(122,96.92){\line(1,0){4}}
\put(129.5,96.92){\line(1,0){4}}
\put(137,96.92){\line(1,0){4}}
\put(144.5,96.92){\line(1,0){4}}

\put(122,104.62){\line(1,0){9}}
\put(134,104.62){\line(1,0){2.2}}
\put(139.2,104.62){\line(1,0){9}}

\put(122,122.76){\line(1,0){26.5}}
\put(132,82.9){$1P$}

\put(164,120.02){\line(1,0){2}}
\put(168,120.02){\line(1,0){2}}
\put(172,120.02){\line(1,0){2}}
\put(176,120.02){\line(1,0){2}}
\put(180,120.02){\line(1,0){2}}
\put(184,120.02){\line(1,0){2}}
\put(188,120.02){\line(1,0){2.3}}

\put(164,124.14){\line(1,0){4}}
\put(171.5,124.14){\line(1,0){4}}
\put(178,124.14){\line(1,0){4}}
\put(185.5,124.14){\line(1,0){4}}

\put(164,132.12){\line(1,0){9}}
\put(176,132.12){\line(1,0){2.2}}
\put(181.2,132.12){\line(1,0){9}}

\put(164,150.98){\line(1,0){26.5}}
\put(177,110.02){$2P$}

\put(135,34){\line(1,0){2}}
\put(139,34){\line(1,0){2}}
\put(143,34){\line(1,0){2}}
\put(147,34){\line(1,0){2}}
\put(151,34){\line(1,0){2}}
\put(155,34){\line(1,0){2}}
\put(159,34){\line(1,0){2.3}}
\put(166,32){$T=0.25$~GeV}

\put(135,24){\line(1,0){4}}
\put(142.5,24){\line(1,0){4}}
\put(150,24){\line(1,0){4}}
\put(157.5,24){\line(1,0){4}}
\put(166,22){$T=0.15$~GeV}

\put(135,14){\line(1,0){9}}
\put(147,14){\line(1,0){2.2}}
\put(152.2,14){\line(1,0){9}}
\put(166,12){$T=0.1$~GeV}

\put(135,4){\line(1,0){26.5}}
\put(166,2){$T=0.0$~GeV}

\put(5,80){$M_{\rm c\bar s}$}
\put(2,70){(GeV)}

\put(30,20){\line(1,0){2}}
\put(20,18){$1.6$}
\put(30,60){\line(1,0){2}}
\put(20,58){$1.8$}
\put(30,100){\line(1,0){4}}
\put(20,98){$2.0$}
\put(30,140){\line(1,0){2}}
\put(20,138){$2.2$}
\end{picture}
\end{center}
\vspace*{8pt}
\caption{The mass spectra of ($c\bar s$)-meson at finite temperature.
\label{fig1}}
\end{figure}
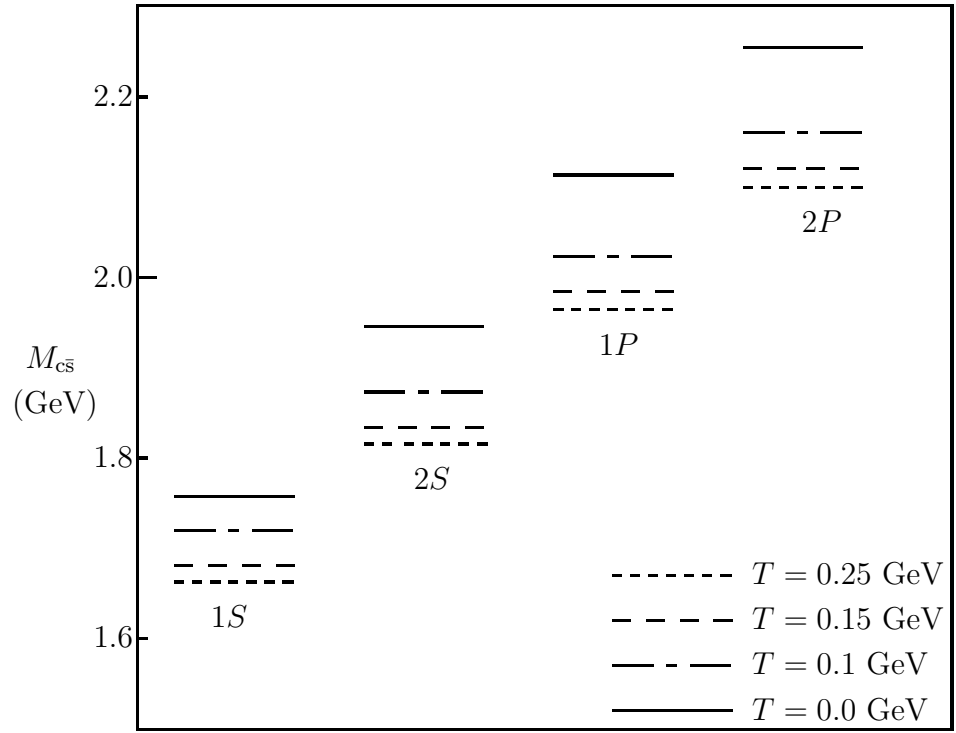

Again, decreasing of the mass by increasing the temperature is clear from these tables.
This effect is shown in Fig.~\ref{fig1} which is the pictoral presentation of some states
for $c\bar s$-system at finite temperature for the values of temperature 0, 0.1, 0.15, 0.25 (in GeV). Results for $(b\bar s)$ and $(b\bar c)$ have a similar behaviour of changing with temperature.

\section{Conclusions}
In this work we have studied spectra of heavy-light mesons at finte temperature.
By combining quark potential model and real-time finite-temperature field theory, the thermofield dynamics Dirac Hamiltonian for heavy-light meson
is presented in the temperature-dependent form. Diagonalizing this Hamiltonian in the non-relativistic harmonic oscillator basis
finite-temperature energy and mass spectra of heavy-light mesons with different heavy and light quark contents are obtained.
It is found that the masses of heavy-light mesons at finite-temperature are less than those at $T=0$.

It is important to remark that TFD has proved to be a
convenient approach for such a study that calculates the
effect of temperature on the spectra of heavy-light quark
mesons~\cite{Ademir}. Furthermore the results will provide a useful input
to studies of quark-gluon plasma at RHIC and LHC.

\section{Acknowledgments}
The authors thank Prof. Dr. Ulrich Mosel for useful comments made by him upon reading
the manuscript of the paper.
This work is supported by the INTAS YS Fellowship (Ref. Nr. 06-1000014-6418) and the grant of the Uzbek Academy of Sciences (FPFI Nr. 41-08).
The work of FCK is supported by NSERCC.

\begin{table}
\centering
\caption{The mass spectra (in GeV) of ($c\bar s$)-meson at finite temperature (in GeV).
The  parameters are chosen as follows:
$m_{\bar s}=0.5$ GeV, $m_{c}=1.486$ GeV, $\lambda=0.8$ GeV$^2$, $V_0=-0.6$ GeV,
$\alpha_s=0.32$.}
\begin{tabular}{ccccc}
\hline
$n(l)/T$&0&0.1&0.15&0.25 \\
\hline
1 (l=0)&1.7575&1.7199&1.6809&1.6623 \\
2&1.9462&1.8730&1.8336&1.8157 \\
3&2.1005&2.0794&2.0323&2.0151 \\
4&2.3061&2.2585&2.2108&2.1968 \\
5&2.4911&2.4693&2.4139&2.3996 \\
6&2.7056&2.6681&2.6124&2.6018 \\
7&2.9113&2.8835&2.8196&2.8095 \\
1 (l=1)&2.1138&2.0231&1.9846&1.9645 \\
2&2.2549&2.1606&2.1207&2.1001 \\
3&2.3950&2.3295&2.2823&2.2626 \\
4&2.5635&2.4896&2.4429&2.4251 \\
5&2.7252&2.6737&2.6196&2.6035 \\
6&2.9122&2.8554&2.8011&2.7869 \\
7&3.0982&3.0517&2.9900&2.9786 \\
1 (l=2)&2.5330&2.4093&2.3732&2.3531 \\
2&2.6414&2.5287&2.4886&2.4663 \\
3&2.7646&2.6673&2.6213&2.6003 \\
4&2.9053&2.8103&2.7629&2.7416 \\
5&3.0498&2.9685&2.9160&2.8976 \\
6&3.2101&3.1318&3.0777&3.0597 \\
7&3.3757&3.3059&3.2470&3.2332 \\
\hline
\end{tabular}
\label{tab1}
\end{table}

\begin{table}
\centering
\caption{The mass spectra (in GeV) of ($b\bar s$)-meson at finite temperature (in GeV).
The parameters are chosen as follows:
$m_{\bar s}=0.5$ GeV, $m_{b}=4.88$ GeV, $\lambda=0.8$ GeV$^2$, $V_0=-0.6$ GeV,
$\alpha_s=0.22$.}
\begin{tabular}{cccccc}
\hline
$n(l)/T$&0&0.1&0.15&0.25 \\
\hline
1 (l=0)&5.1481&4.9984&4.9756&4.9317 \\
2&5.3432&5.1995&5.1741&5.1230 \\
3&5.4986&5.3576&5.3361&5.2764 \\
4&5.7082&5.5695&5.5445&5.4757 \\
5&5.8936&5.7548&5.7343&5.6574 \\
6&6.1098&5.9715&5.9476&5.8603 \\
7&6.3158&6.1767&6.1585&6.0625 \\
1 (l=1)&5.5057&5.3519&5.3269&5.2898 \\
2&5.6503&5.4992&5.4712&5.4251 \\
3&5.7907&5.6414&5.6165&5.5607 \\
4&5.9627&5.8159&5.7877&5.7233 \\
5&6.1253&5.9817&5.9591&5.8858 \\
6&6.3147&6.1718&6.1461&6.0641 \\
7&6.5013&6.3601&6.3400&6.2476 \\
1 (l=2)&5.9259&5.7708&5.7454&5.7147 \\
2&6.0364&5.8820&5.8535&5.8138 \\
3&6.1600&6.0062&5.9780&5.9270 \\
4&6.3028&6.1501&6.1200&6.0609 \\
5&6.4480&6.2986&6.2718&6.2023 \\
6&6.6106&6.4623&6.4348&6.3583 \\
7&6.7771&6.6320&6.6082&6.5203 \\
\hline
\end{tabular}
\label{tab2}
\end{table}

\begin{table}
\centering
\caption{The mass spectra (in GeV) of ($b\bar c$)-meson at finite temperature (in GeV).
The parameters are chosen as follows:
$m_{\bar c}=1.486$ GeV, $m_{b}=4.88$ GeV, $\lambda=0.2$ GeV$^2$, $V_0=-0.3$ GeV,
$\alpha_s=0.22$.}
\begin{tabular}{cccccc}
\hline
$n(l)/T$&0&0.1&0.15&0.25 \\
\hline
1 (l=0)&6.4132&6.0423&5.9721&5.7302 \\
2&6.5940&6.3623&6.1911&5.9703 \\
3&7.0008&6.7803&6.3437&6.3006 \\
4&7.4043&7.1883&6.7436&6.4567 \\
5&7.8131&7.5976&7.1403&6.7371 \\
6&7.9541&7.7572&7.6095&7.1157 \\
7&8.3235&8.1021&7.9986&7.5723 \\
1 (l=1)&7.1444&7.0931&6.4040&5.9279 \\
2&7.3652&7.2417&7.1465&6.4397 \\
3&7.7834&7.3741&7.3373&7.1281 \\
4&7.9654&7.6891&7.6270&7.3249 \\
5&8.2946&7.9966&7.6666&7.6077 \\
6&8.6206&8.3817&7.9385&7.8986 \\
7&9.0263&8.7682&8.3063&7.9818 \\
1 (l=2)&7.3533&7.3150&7.1480&7.0523 \\
2&8.2777&8.2737&8.0734&8.0376 \\
3&9.0333&8.7053&8.6196&8.5776 \\
4&9.2232&8.8580&8.7972&8.7778 \\
5&9.4450&9.1152&9.0458&9.0084 \\
6&9.6959&9.3808&9.3020&9.2559 \\
7&10.0100&9.7292&9.6397&9.5752 \\
\hline
\end{tabular}
\label{tab3}
\end{table}

\newpage
\begin{thebibliography}{0}

\bibitem{babar} B.~Aubert {\it et al.} [BABAR Collaboration], {\it Phys. Rev. Lett.} {\bf 90}, 242001 (2003).
\bibitem{cleo} D.~Besson {\it et al.} [CLEO Collaboration], {\it Phys. Rev. D} {\bf 68}, 032002 (2003).
\bibitem{belle} P.~Krokovny {\it et al.} (Belle Collaboration), {\it Phys. Rev. Lett.} {\bf 91}, 262002 (2003).
\bibitem{qcd1} T.A.~Lahde, C.J.~Nyfalt, D.O.~Riska, {\it Nucl. Phys. A} {\bf 674}, 141 (2000).
\bibitem{godfrey} S.~Godfrey, {\it Phys. Rev. D} {\bf 72}, 054029 (2005).
\bibitem{QCDSF} QCDSF Collaboration, {\it et al.}, {\it Phys. Lett. B} {\bf 652}, 150 (2007).
\bibitem{Narison} Stephan~Narison, {\it Nucl. Phys. B} Suppl. {\bf 152}, 217 (2006).
\bibitem{qcd2} UKQCD Collaboration, J.~Koponen, A.M.~Green, C.~Michael, {\it Nucl. Phys. B} Suppl. {\bf 129-130}, 367 (2004).
\bibitem{ebert} D.~Ebert, V.O.~Galkin and R.N.~Faustov, {\it Phys. Rev. D} {\bf 57}, 5663 (1998).
\bibitem{lattice} R.~Lewis, R.M.~Woloshyn, {\it Nucl. Phys. B} Suppl. {\bf 94}, 359 (2001).
\bibitem{matr1} D.U.~Matrasulov, F.C.~Khanna, Kh.T.~Butanov and Kh.Yu.~Rakhimov, {\it Mod. Phys. Lett. A} {\bf 21}, 1383 (2006).
\bibitem{ndfi04} In {\it Nonlinear Dynamics and Fundamental Interactions}. Eds. F.C.~Khanna and D.U.~Matrasulov. Springer 2006.
\bibitem{rakhimov01} A.~Rakhimov and F.C.~Khanna, {\it Phys. Rev. C} {\bf 64}, 064907 (2001).
\bibitem{dm} D.U.~Matrasulov, F.C.~Khanna and H.~Yusupov, {\it J. Phys. G} {\bf 29}, 475 (2003).
\bibitem{morii} S.N.~Mukherjee {\it et al.}, {\it Phys. Rep.} {\bf 231}, 203 (1993).
\bibitem{tak} Y.~Takahashi and H.~Umezawa, {\it Collective Phenomena} {\bf 2}, 55 (1975). (Reprinted in {\it Int. J. Mod. Phys. A} {\bf 10}, 1755 (1996).)
\bibitem{das} A.~Das, {\it Finite Temperature Field Theory}, (World Scientific, New York, 1997).
\bibitem{Ademir} A.E.~Santana and F.C.~Khanna, {\it Phys. Lett. A} {\bf 203}, 68 (1995).
\bibitem{um} H.~Umezawa, H.~Matsumoto and M.~Tachiki, {\it Thermofield Dynamics}, (North-Holland, Amsterdam, 1982).
\bibitem{tadic} B.~Tadic, R.~Pirc, R.~Blinc, {\it Physica B} {\bf 168},  85 (1990).
\bibitem{egor} B.V.~Egorov, {\it J. Phys.: Condens. Matter} {\bf 4}, 4115 (1992).
\bibitem{matr} D.U.~Matrasulov, F.C.~Khanna, U.R.~Salomov and A.E.~Santana {\it Eur. Phys. J. C} {\bf 42}, 148 (2005).
\bibitem{barnet} S.M.~Barnett and P.L.~Knight, {\it J. Opt. Soc. Am. B} {\bf 2}, 467 (1985).
\bibitem{chai} Chai~Hong and Masahuri~Iwasaki, {\it Int. J. Mod. Phys. A} {\bf 18}, 1497 (2003).
\end{thebibliography}
\end{document}